\documentclass{aa}

\usepackage{graphicx,amssymb,natbib,rotating,psfrag,amsmath,epsfig,array,txfonts}
\bibpunct{(}{)}{;}{a}{}{,} 
\begin{document}

\title{INTEGRAL and XMM--Newton Observations of GRB\,040106\thanks{Based on observations 
with INTEGRAL, an ESA project with instruments and science data centre funded by ESA 
member states (especially the PI countries: Denmark, France, Germany, Italy, 
Switzerland, Spain), Czech Republic and Poland, and with the participation of 
Russia and the USA.}
}

\author{L.\,Moran \inst{1,2}\and
        S.\,Mereghetti \inst{3}\and
        D.\,G\"{o}tz \inst{3} \and
    L.\,Hanlon \inst{1} \and
    A.\,von\,Kienlin \inst{4} \and
        B.\,McBreen \inst{1} \and
    A.\,Tiengo\inst{3} \and
    R.\,Preece\inst{5} \and
    O.R.\,Williams \inst{6} \and
    K.\,Bennett \inst{6} \and
    R.\,M.\,Kippen \inst{7} \and
    S.\,McBreen \inst{1} \and
    S.\,McGlynn \inst{1}
}

\institute{Department of Experimental Physics, University College
  Dublin, Dublin 4, Ireland \and School of Physics and Astronomy, University of Southampton, Southampton, UK \email{lmoran@astro.soton.ac.uk} \and Istituto di Astrofisica Spaziale e
  Fisica Cosmica - CNR, via
  Bassini 15, 20133 Milano, Italy   \and  Max-Planck-Institut
  f\"{u}r extraterrestrische Physik, 85748 Garching, Germany \and Department of Physics, University of Alabama at
 Huntsville, USA \and Science Operations and Data Systems Division of
   ESA/ESTEC,SCI-SDG, NL-2200 AG Noordwijk, The Netherlands \and Space and Remote Sensing Sciences, Los Alamos National
 Laboratory, USA
}

\date{Received 12 May 2004 / Accepted 15 November 2004}

\abstract{On January 6$^{\rm th}$ 2004, the IBAS burst alert system
    triggered the 8$^{\rm th}$ gamma--ray burst (GRB) to be located by the
    \textit{INTEGRAL} satellite. The position was determined
    and publicly distributed within 12\,s, prompting ESA's \textit{XMM--Newton}
    to execute a ToO observation just 5\,hours later, during which an
    X--ray afterglow was detected. The GRB had a duration $\sim$\,52\,s with
    two distinct pulses separated by $\sim$\,42\,s. 
Here we present the results of imaging and spectral analyses of the prompt
    emission from \textit{INTEGRAL} data and the X--ray afterglow from
    \textit{XMM--Newton} data. The $\gamma$--ray spectrum is consistent with a
    single power--law of photon index -1.72\,$\pm$\,0.15.  
The fluence (20\,-\,200\,keV) was 8.2\,$\times$\,10$^{-7}$\,erg\,cm$^{-2}$. 
The X--ray afterglow 
($F_{\nu}(t) \propto {\nu}^{-{\beta_X}} {t}^{-{\delta}}$) 
was extremely hard with $\beta_X = 
    0.47 \pm 0.01$ and $\delta = 1.46 \pm
    0.04$. The 2\,-\,10\,keV flux 11 hours after the burst was 
    1.1\,$\times$\,10$^{-12}$\,erg\,cm$^{-2}$\,s$^{-1}$. 
The time profile of the GRB is consistent with
    the observed trends from previous analysis of BATSE GRBs. 
We find that the X--ray data are not well-fit by either a simple spherical 
fireball or by a speading jet, expanding into a homogeneous medium or 
a wind environment. Based on previously determined correlations between 
GRB spectra and redshift, we estimate a redshift of  
$\sim$\,0.9$^{+0.5}_{-0.4}$ (1\,$\sigma$) and a lower limit on the 
isotropic radiated energy of $\sim$\,5\,$\times$\,10$^{51}$\,erg in this burst.
}

\maketitle

\section{Introduction}
Gamma--ray bursts are an amazingly energetic phenomenon, capable of a
jetted output of order 10$^{51}$\,erg in a few seconds. Although first
detected in the late 1960s, significant progress has mostly been
achieved in the last dozen years. That GRBs are extragalactic in
origin was suggested by the isotropic distribution of GRBs observed by
BATSE on board the \textit{Compton Gamma--Ray Observatory}
\citep{mee1992,fish1994}. The discovery by \textit{BeppoSAX} of
afterglows in the X--ray \citep{costa:1997} and subsequent discoveries
at optical \citep{vanp:1997} and radio \citep{frail:1997} wavelengths
have led to redshift measurements \citep{metz:1997} for $\sim$\,40
bursts ranging from 0.105 to 4.5.

ESA's International Gamma--Ray Astrophysics Laboratory
\textit{INTEGRAL} \citep{wink2003}, launched in October 2002, is composed of
two main telescopes, an imager IBIS \citep{uber2003}, and a
spectrometer SPI \citep{ved2003}, coupled with two monitors, one in the
X--ray band and the other working at optical
wavelengths. Although not built as a GRB-oriented mission,
\textit{INTEGRAL} has a burst alert system, IBAS \citep{mgbwp03}. 
IBAS carries out rapid localisations
for GRBs incident on the IBIS detector with a precision of a few
arcminutes \citep{mere2004a}. The public distribution of these
co-ordinates enables multi-wavelength searches for afterglows at lower
energies. \textit{INTEGRAL} data on the prompt emission in combination
with the early multi-wavelength studies offer the best currently
available probe of these energetic phenomena.

\section{INTEGRAL Observations and Results}
GRB\,040106 was detected by IBAS at 17:55:11\,UTC on 
January\,6$^{\rm th}$ 2004 \citep{mere2004b} 
 with a signal to noise ratio of 7.7 in the 15 to 
200\,keV band for the interval from 17:55:10\,UTC to 17:55:65\,UTC. 
The burst was observed at
$\alpha_{J2000}$\,=\,11$^h$ 52$^m$ 17.7$^s$,
$\delta_{J2000}$\,=\,-46$\degr$ 47\arcmin\,15\arcsec, at an off-axis
angle of 10.5\degr, lying in the partially coded FoV of IBIS and of
SPI, but outside the FoV of the two monitoring instruments, JEM--X and
the OMC. The IBAS alert was automatically distributed approximately
12\,s after the burst start time with a positional uncertainty of only
3\arcmin. Due to the weakness of the burst, data in the energy range
20\,-\,60\,keV from the two time intervals around the prominent peaks
of emission were combined to enable SPI to determine a position for
the GRB. The position for GRB\,040106 extracted from the SPI data is
$\alpha_{J2000}$\,=\,11$^h$\,52$^m$\,51$^s$,
$\delta_{J2000}$\,=\,-46\degr\,47\arcmin\,13\arcsec\,(S/N = 6.8),
 which is 5.7$'$ away from the IBIS location, consistent with the 
10$'$ location accuracy of SPI for a source with S/N $\sim$ 10 \citep{dub+04}. 
At 23:05\,UT \textit{XMM--Newton}
began a 45\,ks exposure ToO observation. A bright source was visible
at $\alpha_{J2000}$\,=\,11$^h$\,52$^m$\,12.4$^s$,
$\delta_{J2000}$\,=\,-46\degr\,47\arcmin\,15.9\arcsec\, in the 30\,ks
Quick-Look-Analysis \citep{ehle2004}, 0.9\arcmin\, from the IBAS
position \citep{tedds2004}.

\subsection{Light Curves}
\begin{figure}
  \begin{center}
\resizebox{\columnwidth}{!}{\includegraphics{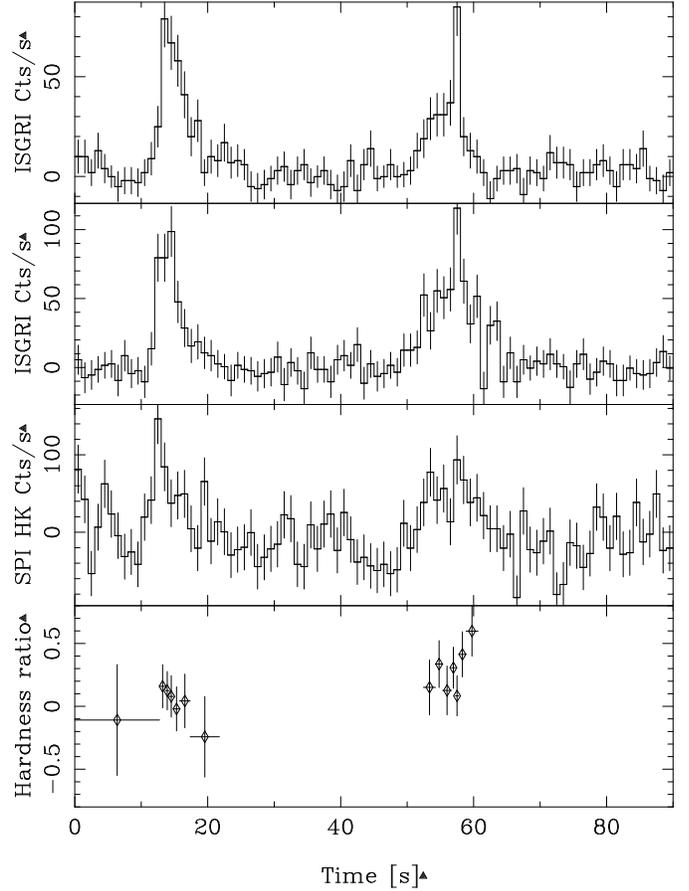}}
    \caption{Light curves of GRB\,040106 beginning at 17:55\,UTC
    extracted from (top to bottom) IBIS/ISGRI data in the range
    15\,-\,40\,keV, IBIS/ISGRI data in the range 40\,-\,200\,keV, SPI
    scientific housekeeping data in the range 20\,keV\,-\,8\,MeV and
    (bottom panel) the hardness ratio evolution derived from
    IBIS/ISGRI count rates.}\label{hr}
  \end{center}
\end{figure}

This GRB falls into the class of long bursts with a duration of $\sim$\,52\,s and two prominent pulses with a peak to peak separation of 42\,s and a
long quiescent interval of $\sim$\,24\,s between the pulses. Light
curves are available from SPI and IBIS. The ISGRI detector of IBIS
\citep{leb2003}, an array of 128\,$\times$\,128 CdTe crystals
sensitive to lower energy $\gamma$--rays, was used to produce the light
curves of GRB\,040106 in the ranges
15\,-\,40\,keV and 40\,-\,200\,keV (Fig.~\ref{hr}). 
The burst light curves were  extracted using the pixels which had at 
least half of their surface illuminated by the GRB. 
The spectral lag could not be determined with any precision due 
to the weakness of the burst. The best constraint was obtained using a 
cross-correlation function on the background-subtracted 
IBIS data in the energy ranges 
15\,-\,40\,keV and 40\,-\,200\,keV and a time-binning of 0.2\,s from 10-65\,s  
after 17:55:00\,UT. The measured lag was $\sim$\,0\,$\pm$\,1\,s which is 
about what can be estimated by eye and does not allow us 
to constrain the GRB to a location on the lag-luminosity diagram, since 
lags are expressed on a logarithmic scale.

The determination of SPI light curves with a short time binning is
only possible by using the 1\,s count rates of the 19 germanium detectors,
which are usually used for scientific house-keeping purposes. These
values reflect the count rates of each detector in the broad SPI
energy band from 20\,keV\,-\,8\,MeV. The SPI light curve in
Fig.~\ref{hr} was generated by summing the background--subtracted count
rates. The background ($\sim$\,50\,counts\,s$^{-1}$\,detector$^{-1}$)
was determined from a 40\,minute period before the burst trigger  
and was subtracted for each detector individually.

A hardness ratio was derived from the ISGRI data by comparing the
count rate in the different energy bands such that HR\,=\,(H\,-\,S)/(H\,+\,S). The
time bins were chosen to ensure at least 100 counts per bin in the sum
of the two bands. The evolution of the hardness ratio is shown in the
bottom panel of Fig.~\ref{hr}. Although visual inspection indicates a
slight hardening in the second pulse, this is not statistically
significant (2\,$\sigma$ level).

The IBIS light curve was denoised using a wavelet analysis
\citep{quil2002} and the temporal properties of the two pulses (rise
time, fall time, FWHM) were extracted and are shown in
Table~\ref{tim}.

\begin{table}
\caption{Temporal properties of the pulses in GRB\,040106.}
\label{tim}
$$
  \begin{array}{p{0.25\linewidth}ccc}
       \hline\\
         \textit{Parameters} & \textit{Rise Time, t$_r$} &
         \textit{Fall Time, t$_f$} & \textit{FWHM}\\
     & \textit{(s)} & \textit{(s)}& \textit{(s)}\\
            \hline\hline
            1$^{\rm st}$ Pulse & 4 & 10 & 5\\
        2$^{\rm nd}$ Pulse & 10 & 6 & 5\\
            \hline
         \end{array}
$$
\end{table}

\subsection{Spectral Analysis}
IBIS is a coded mask instrument and the photons of a single point
source are spread over all the individual detectors. Spectral
extraction is possible using specifically designed software which
consists of modelling the illuminated mask by a point source of
unitary flux placed at the sky coordinates of the GRB. The model is
then fit to the detected shadowgram in each energy channel to
obtain the rate and error for each channel. ISGRI single events were
 used to derive photon indices and fluxes for two time intervals,
of 9\,s and 8\,s duration, around the peaks of emission as shown in
the column `IBIS/ISGRI' of Table~\ref{tspec}. 
Spectral fitting was carried out using two methods. 
Firstly, a time averaged spectrum was derived for GRB\,040106 
by comparing its count rate in different energy bins to the 
corresponding values obtained from the Crab Nebula observed at a 
similar position in the FoV. This method yields a best-fit 
single power--law model with photon index -1.32\,$\pm$\,0.28 
and a fluence of 1.1\,$\times$\,10$^{-6}$\,erg\,cm$^{-2}$. 
Secondly, the effect of the IBIS mask support structure (the so-called
{\it nomex}), which absorbs low--energy ($<$\,50\,keV) photons
differently at different off-axis angles has been 
taken into account using  Offline Software Analysis (OSA) 
version 4.0 off-axis correction matrices and 
the extracted spectra have been modified accordingly.
The improved low-energy response yields a best-fit 
single power--law model with photon index of -1.72\,$\pm$\,0.15 
and a fluence of 8.2\,$\times$\,10$^{-7}$\,erg\,cm$^{-2}$. The 
spectral fit results for the 2 pulses using the {\it nomex} 
correction  are given in Table~\ref{tspec}. Due to the improved
 low energy response, these are the spectral indices used in 
subsequent discussion.  

The spectral evolution of GRB\,040106 was also investigated with
SPI. Spectra were extracted for two time intervals around the peaks of
emission. The background was determined from a 40\,minute period of
SPI data before the burst trigger. The first interval is 7\,s long
starting at the very beginning of the burst at 17:55:11\,UTC, and the
second begins 34\,s later and lasts for 12\,s. Single events
detected by SPI, corrected for intrinsic deadtimes and telemetry gaps,
were binned into 5 equally spaced logarithmic energy bins in the
20\,keV to 500\,MeV range in each of the chosen intervals.  Version 3 
of the OSA from the \textit{INTEGRAL} Science Data Centre 
\citep{couv2003} and the software package SPIROS 6 
\citep{skin2003} were used for SPI spectral extraction, while XSPEC\,11.2
was used for model fitting. The results are shown in the column
labelled `SPI' in Table~\ref{tspec}. The same analysis was conducted 
for multiple events incident on the SPI detectors, 
but yielded no significant improvement to the fit.

\begin{table}
\caption{Spectral analysis results for GRB\,040106 with SPI and IBIS/ISGRI for two
  intervals around the prominent peaks of emission. Errors quoted are for 1
 parameter of interest at 1$\sigma$ confidence level. 
Fluxes are quoted for the energy range 20\,-\,200\,keV. Within the large 
errors on the spectral fits, results are consistent between instruments. }
\label{tspec}
$$
       \begin{array}{p{0.2\linewidth}lccc}
       \hline
       \textit{Interval} & \textit{Parameter} & \mathrm{SPI} & \mathrm{IBIS\,/\,ISGRI}\\
       \hline\hline\\
       1$^{\rm st}$ pulse & \mathrm{Photon~Index} & -1.47^{+0.52}_{-0.59} & -1.70\pm0.12\\
  (9\,s)     & \mathrm{Flux~(erg\,cm^{-2}\,s^{-1})} & 5.9\times10^{-8} & 2.5\times10^{-8}\\
\hline\\
       2$^{\rm nd}$ pulse & \mathrm{Photon~Index} & -1.32^{+0.31}_{-0.34} & -1.41\pm0.1\\
  (8\,s)    & \mathrm{Flux~(erg\,cm^{-2}\,s^{-1})} &  6.1\times10^{-8} & 3.2\times10^{-8}\\
\hline
       \end{array}
$$
\end{table}

\section{XMM--Newton}
\label{xmm}
\textit{XMM--Newton} observed the position of GRB\,040106 for about
45\,ks, starting only $\sim$\,20\,ks after the burst. For the MOS1
camera \citep{turn2001} of the EPIC instrument the medium optical
blocking filter was used, while for the MOS2 and PN detectors
\citep{strud2001} the thin filter was chosen. All three X--ray cameras
operated in Full Frame Mode. The data were processed using SAS version 5.4.1.

\begin{figure}
\begin{center}
      \hspace{0cm}\psfig{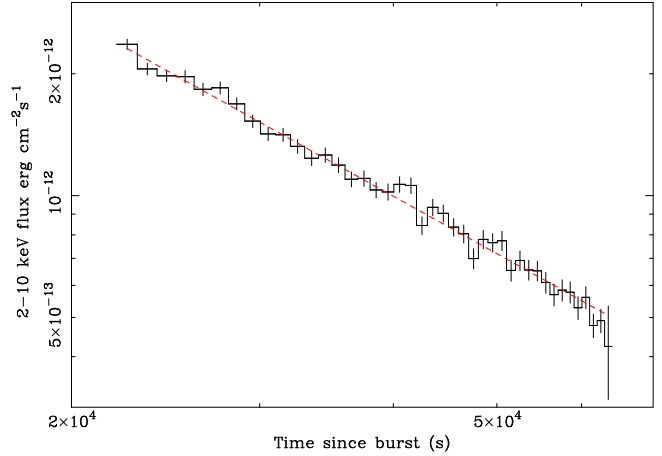}
      \caption{Background subtracted EPIC light curve of the
    X--ray afterglow of GRB 040106 in the 0.4\,-\,10\,keV
    energy band. The dashed line shows the best--fit power--law
    decay $\delta$\,=\,1.46.}
        \label{epic_lc}
\end{center}
\end{figure}

A fading X--ray source was clearly detected in the IBAS error region. The light
curve, obtained by summing the counts from the three EPIC cameras, is
shown in Fig.~\ref{epic_lc}. The afterglow time decay 
($F_{\nu}(t) \propto {t}^{-{\delta}}$) is well fit by a
power--law with index $\delta$\,=\,1.46\,$\pm$\,0.04 (1\,$\sigma$).  
An analysis of the light curves in different energy ranges and of the corresponding
hardness ratios showed no evidence for spectral changes 
during the \textit{XMM--Newton} observation.

\begin{figure}
\begin{center}
      \hspace{0cm}\psfig{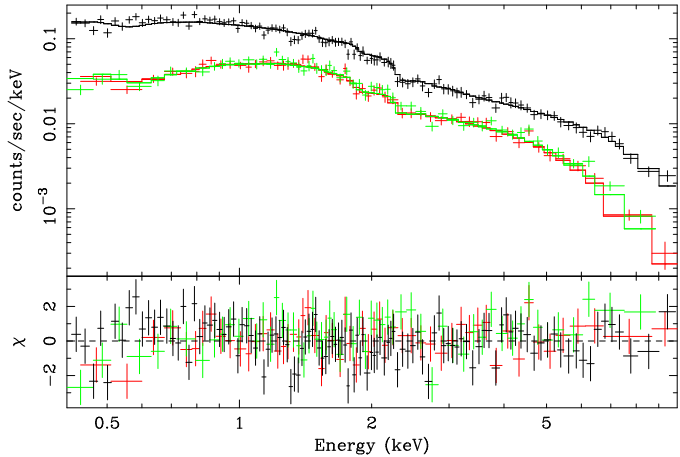}
      \caption{EPIC spectrum of the GRB\,040106 afterglow and its
    best--fit absorbed power--law model. The upper data points
    refer to PN while the lower ones to MOS1 and MOS2. The
    residuals are in units of standard deviations.}
         \label{spec}
\end{center}
   \end{figure}

Afterglow spectra from the PN and MOS cameras were obtained after standard
data screening. The GRB afterglow position was close to a gap between
two chips of the PN detector, so the PN spectra were extracted both from
a circle of 40\arcsec\, radius and from a smaller circle of 20\arcsec\,
radius, which did not contain the gap. Since the two extraction regions led to
consistent results, the spectrum extracted from the larger region,
containing about 90\,\% of the source counts, was used for subsequent
analysis. The background spectra were taken from source free regions of the same
observation. The three spectra in the 0.4\,-\,10\,keV range,
rebinned to have at least 30 counts per bin, are well fit by a power--law model
 (Fig.~\ref{spec}) with spectral index  $\beta_X$\,=\,0.47\,$\pm$\,0.01 
(1\,$\sigma$) (where $F_{\nu} \propto {\nu}^{-\beta_X}$), absorption
N$_H$\,=\,7.4\,$\pm$\,0.9\,$\times$\,10$^{20}$ cm$^{-2}$ and flux of 
1.1\,$\times$\,10$^{-12}$\,erg\,cm$^{-2}$\,s$^{-1}$ (2\,-\,10\,keV) at
11 hours after the burst (observed flux not corrected for absorption).
The column density is consistent with the Galactic
value in this direction ($\sim$8.6\,$\times$\,10$^{20}$ cm$^{-2}$).

From the analysis of the combined MOS and PN spectra of the entire
observation, a 3\,$\sigma$ upper limit of 40 and 200\,eV can be set on
the equivalent width for narrow emission lines in the 0.5\,-\,2 and
2\,-\,8\,keV energy ranges, respectively.

\section{Discussion}
\subsection{Interpretation of the X--ray Afterglow Behaviour}
The X--ray afterglow of GRB\,040106 has a remarkably hard $\beta_X$, which 
is flatter than that of any other X--ray afterglow seen by
\textit{XMM--Newton} \citep{piro2004}. \textit{BeppoSAX} measured  
$\beta_X$ values in this range, but not with such high precision 
\citep{frontera03}. 
GRB\,020405 was observed by Chandra to have $\beta_X=0.72\pm0.21$ 
and $\delta=1.87\pm1.1$ \citep{mph03} compatible, within the large errors, 
with the values measured by \textit{XMM--Newton} in GRB\,040106. 

Following the approach of \citet{pbr+02} we consider the 
suitability of three afterglow models 
using the closure relation $\delta + b\beta_X + c = 0$ where the specific 
values of $b$ and $c$ required for closure depend on the model and whether 
the X--ray frequency, $\nu_{\rm X}$, is above or below $\nu_c$, the cooling frequency of the 
electrons. The models considered are (i) a simple spherical fireball 
expanding into a homogeneous medium (`ISM') \citep{spn98}; 
(ii) a spherical fireball expanding into a wind environment (`Wind') 
\citep{cl99} and (iii) a spreading jet expanding into either a wind-stratified 
medium or the ISM \citep{sph99}.
  
\begin{table}
\caption{Afterglow model tests using X--ray data. The condition for closure 
depends on whether $\nu_c > \nu_X$ or $\nu_c < \nu_X$ (Column 2). $p$ is the power--law 
index of the emitting electron population. }
\label{afterglow}
$$
       \begin{array}{p{0.15\linewidth}lcccc}
       \hline
       \textit{Model} & \mathrm{\nu_c} & b, c & \textit{Closure} & p \\
       \hline\hline\\
    ISM  & \mathrm{> \nu_X}  &  (-3/2, 0)  &  \mathrm{0.76\pm0.06}  & 1.94\pm0.02  \\
    ISM  & \mathrm{< \nu_X}  & (-3/2, 1/2)  &  \mathrm{1.26\pm0.05} & 0.94\pm0.02  \\
\hline
   Wind  & \mathrm{> \nu_X}  &  (-3/2, -1/2)  & \mathrm{0.26\pm0.05} & 1.94\pm0.02 \\
  Wind  & \mathrm{< \nu_X}  & (-3/2, 1/2)  &  \mathrm{1.26\pm0.05}  & 0.94\pm0.02 \\
\hline
  Jet  & \mathrm{> \nu_X}  &  (-2, -1)   & \mathrm{-0.48\pm0.06} &  1.94\pm0.02 \\
  Jet  & \mathrm{< \nu_X}  &  (-2, 0)  & \mathrm{0.52\pm0.06} & 0.94\pm0.02 \\
\hline
       \end{array}
$$
\end{table}

The combination of very hard $\beta_X$ and relatively typical $\delta$ makes it difficult to
fit the data with any of these models (Table~\ref{afterglow}). 
The model which comes closest  to meeting the closure condition is
the case of spherical expansion into a stellar-wind density profile when 
$\nu_c > \nu_X$. This interpretation of the \textit{XMM--Newton}  
data has already been considered in the case of GRB\,040106 \citep{gpp04}.
The results of the \textit{XMM--Newton} analysis presented in Section~\ref{xmm} are consistent 
with those obtained by \citet{gpp04}. 
Our improved constraint on $\delta$ ($\pm$\,0.04 rather than $\pm$\,0.1) 
comes from our use of both PN and MOS camera data, improving the 
light curve statistics and resulting in a less convincing agreement 
(5\,$\sigma$ away from the closure condition) with the wind model. 
The wind interpretation also requires that $\nu_c$  is already above   
$\nu_X$ at the time of the X--ray measurement since no deviation from 
a power--law decay is seen in the \textit{XMM--Newton} lightcurve. 
The required high cooling frequency suggests that the stellar wind has a low
density and/or a low energy fraction for magnetic fields in the blast wave, 
similar to the case of GRB\,020405 \citep{clf04}.  

There are two measurements in R--band (m$_{\rm R}=22.4\pm 0.1$ and $23.7\pm 0.3$) 
of the likely optical afterglow  \citep{mase2004}, 
the first of which was made during the \textit{XMM--Newton} observation, the second 24 hours later. 
The optical flux is consistent with the 
extrapolation of the X--ray spectrum \citep{gpp04}. The
observed temporal decay of the R-band light curve has a power--law
index of 1.2\,$\pm$\,0.4 which is compatible with that seen at
X--ray wavelengths. Therefore, the cooling frequency 
could be below optical wavelengths or above the X--ray 
at the time of the \textit{XMM--Newton} observation. 
\citet{gpp04} argue that the cooling frequency has moved above 
the X--ray band 11 hours post-burst since their \textit{XMM--Newton}
analysis gives good agreement with the wind model prediction for this case. 
We have shown that 
there is no strongly compelling case in favour of the wind model from the 
X--ray data (Table~\ref{afterglow}). 

In fact, the X--ray spectral index is exactly coincident 
with the $1/2$ spectral index `fast cooling' case of the 
synchrotron shock afterglow model
\citep{spn98,zhang2003,piran2000}, 
for the limited spectral region where 
$\nu_X > \nu_c$ (where $\nu_c$ is the synchrotron cooling
frequency) and $\nu_X < \nu_m$, where $\nu_m$ represents the
characteristic synchrotron frequency of the lower-energy limit of
the emitting electron distribution. This limit exists so that
electron energy distributions with a power--law index $p > 2$ do
not diverge at low energies).  
The predicted temporal decay in this case is much flatter than that 
observed, however a more complicated geometry, such as in the case 
of an anisotropic jet, where the emission or bulk Lorentz factor
varies with angle, leaves the time decay index largely
unconstrained \citep{mrp98}. 

For all cases shown in Table~\ref{afterglow}  the appropriate
power--law index for the electron distribution, $p$, is significantly flatter than 
the nearly universal power--law spectrum for charged particles accelerated 
near ultra-relativistic shocks, which has a slope of 2.2\,-\,2.3 \citep{Achter2001}. 
A model of highly collimated jets with flat electron spectra \citep{dc01}, 
although not meeting the closure condition imposed by 
the X--ray data, occupies the right region of the parameter space 
and may, with suitable modification, provide a good fit to the afterglow data 
in this case. 
\subsection{Comparison with other INTEGRAL bursts}
Eighteen GRBs occurred within the field of view of the main \textit{INTEGRAL} 
instruments up to the end of October 2004. 
One of these events, initially classified as a GRB  
(GRB\,040903) is believed to be an X--ray flash (XRF) or 
a possible Type I X--ray flare from a new 
transient source in the Galactic bulge \citep{gmpm04}. 
In a second case, GRB\,031203, 
modelling of the dust-scattered X--ray echo provided the first 
evidence for a low luminosity, XRF source \citep{vwo+04,whl+04}.
The IBIS spectrum, however, is consistent with a single power--law 
of photon index $-1.63\pm0.06$ \citep{sls04}, typical of \textit{INTEGRAL} 
bursts. This event is the only \textit{INTEGRAL} burst to date for which a 
direct redshift ($z=0.1055\pm0.0001$) measurement has been made \citep{pbc+04}.
 
A single power--law model, with photon index in the range -2.0 to -1.0, provides 
a good fit to the data for the vast majority of \textit{INTEGRAL} bursts in the range 
20\,-\,200\,keV \citep{vonk2003}. 
Only one burst detected by IBIS, GRB\,030131, was best fit by a Band model 
over its whole duration, with break energy, E$_0$, of $70\pm 20$\,keV; 
photon index below the turnover, $\alpha$, of $-1.4\pm0.2$ and photon index above the
turnover, $\beta$, of $-3.0\pm1.0$ \citep{gmh+03}. 
GRB\,030131 occurred during an \textit{INTEGRAL} slew and due to telemetry restrictions  
only limited SPI data were obtained for this event \citep{moranetal04}.

\subsection{Constraints on redshift and energy}
A relationship between $\alpha$, the $\gamma$--ray spectral index below 
the spectral turnover, and redshift, derived from {\it BeppoSAX} $\gamma$--ray 
 bursts with known redshifts, is of the form \citep{aft+02}:
\begin{equation}
|\alpha| = (2.76\pm0.09)(1+z)^{-0.75\pm0.06} 
\label{eq:amati}
\end{equation}
This relationship reflects a dependence of $\alpha$ on the peak energy, 
E$_{\rm p}$, due to spectral curvature, rather than a fundamental connection 
between $z$ and $\alpha$. Assuming the IBIS photon index is $\alpha$, the 
redshift of GRB\,040106 derived from Eq.~\ref{eq:amati} is 
$z \sim 0.9^{+0.5}_{-0.4}$ (1\,$\sigma$). This is a reasonable assumption since 
most $\beta$ values are steeper than -2, while most 
$\alpha$ values are in the range -2 to 0 \citep{pbm+00}. 
We estimate 200\,keV as the lower limit on the break energy, since there is 
no curvature evident in the \textit{INTEGRAL} spectra up to this energy. 
In the cosmological rest-frame, using the lower limit on $z$ from above,
$E_{\rm 0} > 300$\,keV. The peak energy, $E_{\rm p}$, is given by
$(2+\alpha) \times \,E_{\rm 0} $ and hence $E_{\rm p} > 84$\,keV. 
The relationship $E_{\rm p} \propto {E_{\rm rad}}^{0.52} $ \citep{aft+02} 
can then be used to estimate a lower limit on the isotropic radiated energy,  
$E_{\rm rad}$, of $\sim 5\,\times 10^{51}$ erg in this burst. 
Due to the weakness of this GRB, no strong constraint on the spectral lag, 
and hence on the peak luminosity from the luminosity-lag relationship, 
could be determined \citep{nmb00}. 

The lower limit on the isotropic radiated energy is a factor of 10 higher 
than the `mean' $E_{\gamma} \sim 5\times 10^{50}$\,erg determined by 
\citet{fks+01} when geometric corrections due to the effects of a jet are 
taken into account. \citet{liang04} has investigated the relationships between 
jet opening angle, $\theta$, and the prompt $\gamma$ emission and the 
X--ray afterglow emission in a sample of 10 GRBs. The 20\,-\,2000\,keV fluence 
as a function of $\theta$ is found to be a broken power--law, with a 
break at $\theta=0.1$. GRB\,040106 has an estimated 20\,-\,2000\,keV fluence 
of $6.8\times 10^{-8}$\,erg\,cm$^{-2}$. This is probably an over-estimate since it 
assumes the power--law spectrum extends up to 2\,MeV, which is almost certainly 
not the case. However, it allows us to determine a lower limit on $\theta$ from 
the $\gamma$--ray data from the relation: 
\begin{equation}
S_{\gamma,-6} = 0.025 \times {\theta}^{-3.79}
\end{equation} 
for $\theta > 0.1$, where $S_{\gamma,-6}$ is the $\gamma$--ray fluence in units of 
$10^{-6}$\,erg\,cm$^{-2}$ \citep{liang04}. 
A lower limit of $\theta > 0.22$\,rad is obtained from this 
relation. Hence, a lower limit 
of $1.2\times 10^{50}$\,erg on the energy radiated in $\gamma$--rays is derived. 
Uncertainties in the cosmological correction 
 which should be applied, and in the 
true spectral shape,  probably give rise to an uncertainty of order 2 in 
the $\gamma$--ray fluence, which is $\sim 50$\% on $\theta$. 

The $\theta$ value predicted from the X--ray afterglow decay 
 slope of $\delta$ \citep{liang04} according to the relation:
\begin{equation}
\delta = (1.6\pm0.07) - (2.27\pm0.54)\times \theta 
\end{equation}
is $\sim 5\times 10^{-2}$, considerably smaller than 
the lower limit derived from the $\gamma$--ray data. If this correction is applied to the 
isotropic radiated energy estimate, then a lower limit of $6\times 10^{48}$\,erg 
on the energy radiated in $\gamma$--rays is obtained. 

A third relationship, between the rest-frame X--ray afterglow flux 10 hours after 
the burst and $\theta$, is given by:
\begin{equation}
F_{x,-13} = 0.47\times {\theta}^{-1.8\pm0.4}
\end{equation} 
where $F_{x,-13}$ is the rest-frame 2\,-\,10\,keV flux derived from the observed 
flux in units of  $10^{-13}$\,erg\,cm$^{-2}$\,s$^{-1}$  according to:
\begin{equation}
F_x = \frac{F_{x,obs}}{1+z}(1+z)^{\beta_X-1}
\end{equation}

Using $z\sim 0.9$ derived above, a $\theta$ value of 0.24\,rad is obtained, consistent 
with the estimate from the $\gamma$--ray data. 
 
From the burst morphology view-point, GRB\,040106 has the characteristics 
of a low-luminosity, long lag burst (few, well-separated peaks) \citep{norris02}.   
Its location in super-galactic coordinates (161$^{\circ}$, -22$^{\circ}$) is 
consistent with that found in the subset of such 
 bursts \citep{norris02}, but there is no independent 
evidence to suggest that this event is anything other than a `standard' 
cosmological burst. The pulse properties and time intervals between pulses are related to
T$_{90}$ \citep{mmqh02}. The time interval, $\Delta$T, and properties
of the two pulses of GRB\,040106, including rise time, fall time
and FWHM fit well with the expected trends from previous analyses of
BATSE bursts \citep{mmqh02}.

\subsection{GRB-Afterglow Connection}

\begin{figure}
\begin{center}
      \hspace{0cm}\psfig{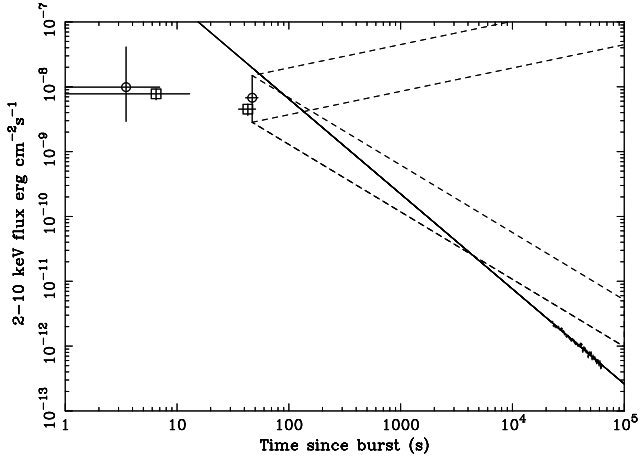}
      \caption[]{Extrapolated 2\,-\,10\,keV fluxes from SPI (circles) and 
IBIS (squares) for each of the two pulses. The
      solid line is the extrapolation of the X--ray data while the
      dashed lines represent the range of decay slopes based on the
      power--law decay of the 2$^{\rm nd}$ pulse.}
         \label{ag}
\end{center}
   \end{figure}

We examined the possibility that the second pulse of the GRB
is the onset of the afterglow. The temporal decay of the 2$^{\rm nd}$ pulse is
consistent with a power--law (with large errors due to the 
small number of bins and their low counts) which may indicate the presence of a
high-energy afterglow, due to external shocks, during the burst
itself. 

The IBIS and SPI derived fluxes in the two pulses were 
extrapolated to X--ray energies, assuming a single power--law spectrum, 
to estimate the 2\,-\,10\,keV flux during the GRB. Clearly
this is valid only if the GRB has neither an X--ray excess nor is X--ray rich.
It is unlikely that this is an X--ray rich GRB with a peak energy at or
below the low end of the SPI and IBIS detector sensitivity
(i.e. $\sim$\,20\,keV) since the photon index would then correspond to an
unusually hard value for the high-energy index above the spectral
turnover \citep{pbm+00}. It is more likely that the weakness of this
GRB washes out evidence for a spectral break at more typical energies
of a few hundred keV.

The extrapolated fluxes from SPI and IBIS are shown as symbols in Fig.~\ref{ag}. 
Folding in the range of possible values of the SPI and IBIS extrapolated fluxes,
 assuming they follow the power--law temporal decay of the 2$^{\rm nd}$ GRB pulse, 
these extrapolated values fall marginally below the backward extrapolation of the
late-time X--ray afterglow data obtained by \textit{XMM--Newton} (Fig.~\ref{ag}).
Assuming the decay slope of the 2$^{\rm nd}$ pulse is indicative
of the onset of a high-energy afterglow, the change in temporal decay
occurs roughly between 50 and 5\,000\,s after the onset of the GRB  
 and may be evidence of the passage of the cooling break through the X--ray band 
in that time window. The fact that the $\gamma$--ray photon index and 
X--ray spectral index are within 2\,$\sigma$ of each other 
suggests that $\nu_c < \nu_X$ during the X--ray observations. This interpretation 
hinges on the assumption that the decay of the 2$^{\rm nd}$ GRB pulse is the onset of the 
afterglow.

There is no evidence in the IBIS/ISGRI light
curve for soft extended or delayed emission such as that observed by,
for example, SIGMA/GRANAT in GRB\,920723 \citep{buren1999} or by
HETE-II in GRB\,021211 \citep{crew2003}.

\section{Conclusion}
\textit{INTEGRAL}'s capabilities for GRB studies, from the
rapid and accurate localisation provided by IBAS, to the good timing
and spectral resolution provided by SPI and IBIS, are well 
illustrated in the case of GRB\,040106. 
 The $\gamma$--ray measurements have provided  
a constraint on the redshift ($z \sim 0.9^{+0.5}_{-0.4}$) 
and energy (lower limit on the isotropic radiated energy 
of $\sim 5\,\times 10^{51}$ erg) for this burst, 
in the absence of optical spectroscopy. Furthermore, previously identified 
correlations between $\gamma$--ray properties and X--ray afterglows have 
been used to estimate the jet opening angle ($\sim 0.22$\,rad) and 
hence the geometry-corrected radiated energy ($1.2\times 10^{50}$\,erg) 
in this burst. 

 The combination of very hard energy index and 
relatively typical temporal decay makes it difficult to 
fit the X--ray data with afterglow models of a simple spherical fireball 
or a jet, expanding into either the ISM or a wind environment. In particular, 
we can rule out the case of isotropic expansion into a wind environment ($\nu_c > \nu_X$), 
suggested by \citet{gpp04}, at 5\,$\sigma$ confidence. In all scenarios, the 
power--law index of the emitting electrons is flatter than 2 and models involving flat 
electron spectra may be more successful fitting the X--ray afterglow data of this 
burst.

The additional science provided by the combination of $\gamma$--ray and 
X--ray early afterglow measurements highlight the
importance of broadband high-energy coverage from the GRB through to
the first days post-burst. We anticipate the significant progress in this 
area that will be made with the launch of NASA's Swift mission.  
\acknowledgements{We thank the anonymous referee for 
his/her useful comments which helped to improve this paper.

\bibliography{moranrefs}
\bibliographystyle{aa}
\end{document}